# IAX-Based Peer-to-Peer VoIP Architecture

**Amor Lazzez, Ouissem Ben Fredj, Thabet Slimani**

**Taif University,**
**Kingdom of Saudi Arabia**

**Abstract**

Nowadays, Voice over IP (VoIP) constitutes a privileged field of service innovation. One benefit of the VoIP technology is that it may be deployed using a centralized or a distributed architecture. The majority of today's VoIP systems are deployed using the client–server centralized architecture. One of the most efficient approaches used in the deployment of centralized VoIP systems is based on the use of IAX (Inter-Asterisk Exchange), an open-source signaling/data exchange protocol. Even though they are currently widely used, client-server VoIP systems suffer from many weaknesses such as the presence of single points of failure, an inefficient resources management, and system non-scalability. In order to help the development of scalable and reliable VoIP systems, the development community starts tending towards the deployment of the VoIP service using a peer-to-peer distributed architecture. The aim of this paper is to develop an IAX-based peer-to-peer VoIP architecture, an optimized VoIP architecture that takes advantage of the benefits of the IAX protocol and those of the peer-to-peer distribution model.

***Keywords:*** *VoIP, Peer-to-peer, IAX, Kademlia*

## 1. Introduction

Voice over IP (VoIP)[1] is a technology for the delivery of voice communications and multimedia sessions over an IP (Internet Protocol) network, such as the Internet. The VoIP technology allows many benefits for customers and communication services providers. In fact, the VoIP approach allows the reduction of calls and communication infrastructure costs, helps the provision of new communication services (instant messages, video calls, images transfer, etc.), ensures users and services mobility, allows the integration and collaboration with other applications (email, web browser, instant messenger, social-networking applications), and provides an online tracking and managing system. These significant benefits are behind the prevalence of VoIP compared to legacy phone systems. Actually, most service providers have started or at least have planned to migrate their PSTN (Public Switched Telephone Network) infrastructure to an IP-based one.

One benefit of the VoIP technology is that it may be deployed using either a centralized or a distributed architecture. The majority of today's VoIP systems are deployed using a client-server centralized architecture [2]. A client-server VoIP system relies on the use of a set of interconnected central servers responsible for the registration of users, and the management of VoIP sessions between registered users.

Different signaling protocols have been proposed for the deployment of client-server VoIP systems such as H.323 [2], SIP [3], and IAX [4]. The current VoIP systems mainly rely on the use of SIP, and IAX signaling protocols. Even though it was proposed for security and flexibility purposes, SIP suffers from many weaknesses [2,3]. In fact, nowadays SIP becomes more and more complex due to the incremental modification of SIP specifications in order to improve the protocol adaptability. Moreover, SIP suffers from the difficulties of crossing NAT (Network Address Translation) and firewall boxes. IAX protocol is considered as a possible candidate to solve SIP problems [5, 6]. In fact, IAX is a simple protocol which supports NAT and firewalls traversal since no IP addresses are enclosed in IAX signaling messages. Moreover, IAX allows signaling and data traffic exchange in contrast with SIP which is limited to the signaling task.

Even though they are currently widely used, client-server VoIP systems suffer from many hurdles. The main issues of the client-server VoIP systems are single points of failure, scalability, service availability, and security. In order to overcome the shortcomings of the client-server model, and help the development of scalable and reliable VoIP systems, the development communities start tending towards the deployment of the VoIP service using a peer-to-peer decentralized architecture. A peer-to-peer VoIP system [7,14] allows service provision through the establishment of a symmetric collaboration between the system nodes (peers) interconnected according to a given logic architecture (overlay). This helps the elimination of the single points of failure, the increase of the system scalability, the enhancement of the system efficiency, the decrease of the system cost, and thus the increase of the system cost-effectiveness. The peer-to-peer model may be deployed using different overlay architectures such as Can, Chord, and Kademlia [8,9,10]. Kademlia peer-to-peer system is considered as the most efficient architecture in the deployment of peer-to-peer VoIP systems [8,9,10].

Given the benefits of the peer-to-peer distribution model, l'IETF has recently started working on the development of a peer-to-peer signaling protocol (P2PSIP [11]) to help a peer-to-peer deployment of the VoIP service based on standardized protocols. P2PSIP allows the use of SIP in environments where the service of establishing and managing sessions is mainly handled by a collection of intelligent end-points, rather than centralized SIP servers. The current P2PSIP scenarios only consider the infrastructure for the connectivity inside a single domain. In [12], the authors propose an extension of the current work to a hierarchical multi-domain scenario: a two level hierarchical peer-to-peer overlay architecture for the interconnection of different P2PSIP domains.

Despite the advantages of the IAX-based scenarios compared to SIP-based scenarios in the deployment of centralized VoIP systems, and in spite of the benefits of the peer-to-peer distribution model, no effort has been made to incorporate the peer-to-peer technology in the deployment of IAX-based VoIP systems. In this paper, we propose an IAX-based peer-to-peer VoIP architecture; an optimized architecture that takes advantage of the benefits of the IAX protocol and those of the peer-to-peer distribution model. The VoIP architecture relies on the use of the Kademlia overlay architecture as the most efficient peer-to-peer distribution model in the deployment of peer-to-peer VoIP systems.

The remaining part of this paper is organized as follows. Section 2 presents a detailed overview about the current VoIP systems. First, we highlight the benefits of the VoIP service. Then, we present the client-server architecture, the main used architecture in the deployment of the current VoIP systems. Next, we present the main used signaling protocols in the deployment of the client-server VoIP architecture such as H.323, SIP, and IAX. Finally, we show that IAX-based scenario is the most efficient approach in the deployment of centralized VoIP systems. In section 3, we show the need to migrate towards a peer-to-peer architecture in the deployment of the VoIP systems. First, we present the limits of the client-server model in the deployment of VoIP systems. Then, we present the peer-to-peer model, and we show how it helps the development of scalable and reliable VoIP systems. Section 4 presents an overview about the peer-to-peer architecture. First, we present the main peer-to-peer architectures such as Can, Chord, and Kademlia. Then, we show that Kademlia is the most efficient architecture in the deployment of peer-to-peer architecture. Section 5 provides a detailed presentation of the proposed VoIP architecture. First, we present the system architecture. Then, we present the proposed protocols for the deployment of the considered architecture. Section 6 concludes the paper.

## 2. VoIP: An Overview About the Current Systems

VoIP is a rapidly growing technology that delivers voice communications over Internet or a private IP network instead of the traditional telephone lines. VoIP involves sending voice information in the form of discrete IP packets sent over Internet rather than an analog signal sent throughout the traditional telephone network.

### 2.1 VoIP Benefits

The use of the VoIP technology allows many benefits for users, companies, and services providers. The key benefits of the VoIP technology are as follows [1, 2]:

**Cost savings:** The most attractive feature of VoIP is its cost-saving potential. Actually, for users, VoIP makes long-distance phone calls inexpensive since telephone calls over the Internet do not incur a surcharge beyond what the user is paying for Internet access. For companies, VoIP reduces cost for equipment, lines, manpower, and maintenance. In fact, thanks to the VoIP technology, all of an organization's voice and data traffic is integrated into one physical network, bypassing the need of two separate networks. For service providers, VoIP allows the use of the same communication infrastructure for the provision of different services (voice and video communications, data transfer, etc.), which reduces the cost of services deployment.

**Provision of new communication services:** The legacy phone system mainly provides voice and fax service. However, the VoIP technology allows the provision of new communication services in addition to the basic communications services (phone, fax). In fact VoIP allows users to check out friends' presence (such as online, offline, busy), send instant messages, make voice or video calls, and transfer images, and so on.

**Phone portability:** with the traditional phone system, a phone number is dedicated to a physical phone line. Hence, a user cannot move his home phone to another place if his wants to use the same phone number. Whereas, VoIP provides number mobility; the phone device can use the same number virtually everywhere as long as it has proper IP connectivity. Many businesspeople today bring their IP phones or soft-phones when traveling, and use the same numbers everywhere.

**Service mobility**: Wherever the user (phone) goes, the same services will be available, such as call features, voicemail access, call logs, security features, service policy, and so on.

**Integration and collaboration with other applications**: VoIP allows the integration and collaboration with other applications such as email, web browser, instant messenger, social-networking applications, and so on. The integration and collaboration create synergy and provide valuable services to the users. Typical examples are voicemail delivery via email, voice call button on an email, and presence information on a contact list.

**User control interface**: Most VoIP service providers provide a user control interface, typically a web GUI, to their customers so that they can change features, options, and services dynamically. For instance, a user may login into the web GUI and change presence information (online, offline), anonymous call block, etc.

## 2.2 Client-Server VoIP Architecture

One of the main features of the VoIP technology is that it may be deployed using a centralized or a distributed architecture. The majority of current VoIP systems are deployed using a client-server centralized architecture. A client-server VoIP system relies on the use of a set of interconnected central servers known as gatekeepers, proxy servers, or soft-switches. The central servers are responsible for users' registration as well as the establishment of VoIP sessions between registered users. Figure 1 shows an example of a VoIP system deployed using the client-server architecture. As it is illustrated in the figure, each central server handles (registers, establishes a session with a local or a distant user, etc.) a set of users. Each user must be registered on one of the central servers (registrar server) to be able to exchange data with other registered users. A user gets access to the service only over the registrar server.

## 2.3 Deployment of a Client-Server VoIP System

The deployment of a client-server VoIP system involves two protocols; a signaling protocol used for the establishment of a VoIP session between two registered users, and a data transmission protocol used for data transmission during an established VoIP session.
The majority of client-server VoIP systems rely on the use of the RTP (Real-Time Transport Protocol) protocol for data transmission during a VoIP session.

Different signaling protocols have been proposed for the deployment of client-server VoIP systems. In earlier stages of telephony over IP (ToIP) deployments, H.323 [2] had started to attract service providers. H.323 was very complex and includes dozens of additional protocol. For need of flexibility and security, most of VoIP companies has adopted SIP [3] as their signaling protocol. Nowadays, SIP becomes more and more complex. The RFC of SIP includes 628 occurrences of 'MUST', 342 of 'SHOULD' and 377 of 'MAY' occurrences. These complications are the result of incremental modification of SIP specifications in order to solve the problems caused by the non-compliance of SO with the OSI model and uses information which belongs to underlying layers. SIP suffers from the difficulties of crossing NAT (Network Address Translation, [NAT]) and firewall boxes. SIP is based on path-decoupled paradigm which means to create two paths for each VoIP connection. This architecture leads to insert an intermediate node in both the signaling and the media path for access-control and billing purposes.

IAX (Inter-Asterisk Exchange, [4, 5, 6]) protocol is a possible candidate to solve SIP problems. The main features of IAX are:

- IAX uses UDP (User Datagram Protocol) with a single port number 4569.
- The IAX registration mechanism is similar to SIP; an IAX registrant contacts a registrar server with specific messages.
- IAX couples signaling and media paths in contrast with the path-decoupled approach adopted by SIP. However, IAX allows decoupling once the connection has been successfully established.
- IAX does not require a new protocol for the exchange of media streams. It handles media streams itself. Various social media types may be sent by IAX: voice, video, image, text, HTML. A multiplexing capability is supported by IAX to distinguish ongoing sessions using two application layer identifiers: Source Call Number and Destination Call Number.
- IAX uses unreliable messages for media and reliable messages for control messages.
- IAX supports NAT traversal since no IP addresses are enclosed in IAX signaling messages.
- IAX defines a set of messages used to monitor the status of the network. These messages can be exchanged during or outside an active call.
- IAX allows exchange of shared keys. It may be used either with plain text or in conjunction with encryption mechanisms like AES (Advanced Encryption Standard). Unlike SIP, no confusion is raised by identity related information used to enforce authentication. Also, IAX exchange authentication requests which enclose a

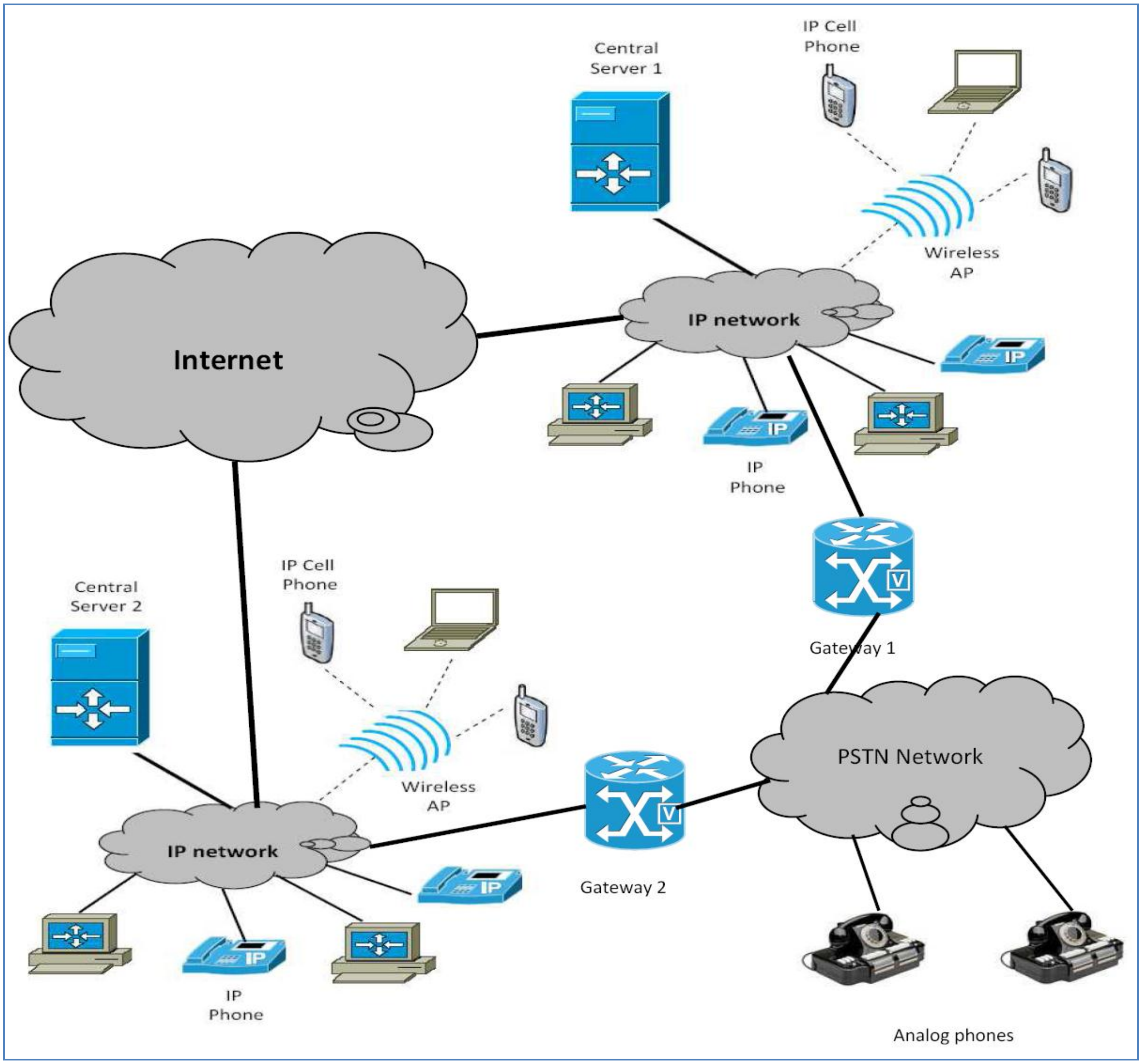

Figure1: Client-Server VoIP Architecture: An illustrative Example

security challenge according to the encryption method used.

- IAX can be easily deployed to provide heterogeneous calls between IPv4 and IPv6-based peers.

## 3. Towards Peer-To-Peer VoIP Systems

One benefit of the VoIP technology is that it may be deployed using either a client-server centralized architecture or a peer-to-peer distributed architecture.

The majority of today's VoIP systems are deployed according to the client-server model. A client-server VoIP system relies on the use of a set of interconnected central proxies responsible for the registration of users, and the management of VoIP sessions between registered users.

Even though they are currently widely used, client-server VoIP systems suffer from many hurdles. The main issues of the client-server VoIP systems are single points failure, scalability, availability, and security [13].

**Scalability issue:** with the client-server VoIP systems, the users under a given central proxy share the available resources on such server. Thus, as more users join the system, fewer resources are available to serve each user, and hence a slower data transfer for all users. Therefore, large investments in system infrastructure (redundant components, advanced monitoring applications, etc.) will be usually needed each time the number of system users is increased. This results in a scalability problem especially with the ever increasing of the users of the VoIP service.

**Single points of failure:** With the client-server VoIP architecture, a user can only be served by the central proxy where it is registered (registrar proxy). Thus, the failure of a central proxy results in the failure of all the users registered on such proxy to establish a VoIP session. Hence, a central proxy in a client-server VoIP system represents a single point of failure for all users registered on such proxy.

**Availability issue:** As it is mentioned above, with the client-server VoIP architecture, a user can only be served by the registrar proxy. Thus, the failure of a central proxy results to the unavailability of the VoIP service for all users registered on such proxy.

**Security issue:** A client-server VoIP system may be vulnerable to denial-of-service (DoS) security attacks due to the presence of a single point of failure. In fact, the failure of a central proxy due to a security attack leads to the unavailability of the VoIP service for all users registered on such proxy; thus, a DoS attack of the VoIP system.

In order to overcome the shortcomings of the client-server model, and help the development of scalable and reliable VoIP systems, the development communities start tending towards the deployment of the VoIP service using a peer-to-peer decentralized architecture. Actually, a peer-to-peer VoIP system [7, 14] allows service provision through the establishment of a symmetric collaboration between the system nodes (peers) communicating according to a given logic architecture (overlay). This helps the guarantee of the following advantages:

**System without single points of failure:** with the peer-to-peer VoIP system, a user gets access to the service through one of the available system peers and not through a specific central server. This results to the fact that, the failure of one or multiple nodes in a peer-to-peer system will note results in service unavailability for any users. Thus, the elimination of single points of failure thanks to the distribution of system's resources.

**More reliable system:** the distribution nature of the peer-to-peer systems allows the elimination of single points of failure. This makes the system more available and not vulnerable to any DoS security attack. In fact, if one of the system peer fails to function properly due to any reason (security attack or other), the whole system will not be compromised or damaged, and thus, the service will be available for all system users.

**More scalable system:** the peer-to-peer system may support the increase of the number of users without the need of any investment in the system infrastructure like with client-server systems. In fact, after being connected to a peer-to-peer system, a user node makes its available resources (bandwidth, storage space, and computing power) at the disposal of the system. This increases the capacity of the system, and thus enhances its scalability. Unlike with client-server systems, with peer-to-peer systems, the increase of the number of the connected users increases the available resources for each user which leads to a faster data transfer for all users.

**Higher cost-effectiveness:** any node in a peer-to-peer system, can act as both a server and a user workstation. Therefore, a peer-to-peer VoIP system may be deployed without the use of dedicated servers with special hardware and software configurations. These results in the reduction of the system cost compared to the client-server system which relies on the use of special servers. As it is shown above, the peer-to-peer model allows better performance compared to the client-server model. For instance, the increase of the users number, leads to better performances with peer-to-peer systems, and worse performances with client-server systems. Therefore, the peer-to-peer model ensures better performances with less cost compared to client-server model.

**Efficient system resources utilization:** the peer-to-peer model ensures more efficient resources utilization compared to the client-server model. In fact, with the peer-to-peer model, system resources are distributed over all the system nodes, and shared among all the connected users. Hence, a user gets access to the service if the needed resources are available on one of the system nodes. However, with the client-server model, a user may not get access due to the lack of resources in registrar proxy while resources are available in the other central servers.

Given the benefits of the peer-to-peer distribution model, l'IETF has recently started working on the development of P2PSIP [11], a peer-to-peer signaling protocol allowing a peer-to-peer deployment of the VoIP service based on standardized protocols. P2PSIP re-implements the functionalities of SIP (users' registration and localization, signaling traffic routing, etc.) in a decentralized fashion. The user and service information are distributed among all peers in the peer-to-peer overlay network, instead of storing it in the registrar and proxy servers. The requests for this information are also handled by the overlay infrastructure. The advantages of P2PSIP include the elimination of the single points of failure and the reducing of the systems ‘costs as it does not require any dedicated equipment.

The P2PSIP scenarios currently proposed by the IETF only consider the infrastructure for the connectivity inside a single domain. In [12], the authors propose an extension of the current work to a hierarchical multi-domain scenario: a two level hierarchical peer-to-peer overlay architecture for the interconnection of different P2PSIP domains.

The peer-to-peer distribution model may be deployed using different overlay architectures. The aim of the following section is to present a detailed overview about the peer-to-peer distribution model, as well as the main considered overlay architectures such as Can, Chord, and Kademlia.

## 4. Peer-to-Peer distributed architectures

As stated above, the peer-to-peer architecture seems to be a good alternative for VoIP traffic. Peer-to-peer overlay networks do not have any hierarchical organization or centralized control. Peers form self-organizing overlay networks providing robust routing architecture, efficient search of data items, redundant storage (high availability), massive scalability, and fault tolerance.

Peer-to-peer systems are categorized into unstructured and structured system. In unstructured peer-to-peer system, the placement of content is completely unrelated to the overlay topology which does not impose any structure on the overlay networks. In structured networks the overlay topology is controlled and information is placed at precisely specified locations which lead to overlays with specific topologies and properties. The structured peer-to-peer systems insure a high control of the network and its resources and provide a stable and load balanced architecture. Thus, we adopt the structured peer-to-peer architecture for the proposed VoIP system.

Structured peer-to-peer systems use the Distributed Hash Table (DHT) as a substrate in which objects (or data, identified by keys) provided by a peer (identified by unique NodeID). Keys are mapped by the overlay network protocol to a unique live peer in the overlay network. The peer-to-peer overlay networks insure a scalable storage and retrieval of {key,value} pairs on the overlay network. Given a key, a store operation (put(key,value)) or a lookup retrieval operation (value=get(key)) can be invoked to store and retrieve the data object corresponding to the key, which involves routing requests to the peer corresponding to the key.

Each peer maintains a small routing table of its neighboring peers (NodeIDs and IP). Lookup queries or message routing are forwarded across overlay paths to peers with NodeIDs that are closer to the key in the identifier space. The DHT guarantees a complexity of the routing request around a small O(logN) overlay hops, where N is the number of peers in the system.

Table 1summarizes the performance of five existing structured peer-to-peer systems CAN [9], CHORD [8], TAPESTRY [15], PASTRY [16], and KADEMLIA [10]:

The performance analysis and the complexity of the overlay algorithm show that Kademlia presents a tradeoff between robustness, routing performance, and complexity of the algorithms. Moreover, Kademlia is the most used DHT in real applications. For instance, Overnet network [17] is used in MLDonkey. Also, Kad Network [18] is used by eMule8, aMule, RevConnect10, BitTorrentAzureus DHT, BitTorrent Mainline DHT, µTorrent, BitSpirit, Bit-Comet, and KTorrent.

Kademlia uses a XOR metric for distance between points in the key space. XOR is symmetric, allowing peers to receive lookup queries from precisely the same distribution of nodes contained in their routing tables. Kademlia can send a query to any node within an interval, allowing it to select routes based on latency or even send parallel, asynchronous.

XOR metric measures the distance between two IDs by interpreting the result of the bit-wise exclusive OR function on the two IDs as integers. For example, the distance between the identifiers 4 and 7 is 3. Considering the shortest unique prefix of a node identifier, the metric treats the nodes and their identifiers as the leaves of a binary tree. For each node, Kademlia further divides the tree into sub-trees not containing the node, see figure 2.

With its XOR metric, Kademlia's routing has been formally proved consistent and achieves a lookup latency of O(log(N)). The required amount of node state grows with the size of a Kademlia network. However, it is configurable and together with the adjustable parallelism factor allows for a trade-off of node state, bandwidth consumption, and lookup latency.

The peer in the network stores a list of {IP address, UDP port, NodeID} triples for peers of distance between 2i and 2i+1 from itself. These lists are called k-buckets. The value

| | CAN | Chord | Tapestry | Pastry | Kademlia |
|---|---|---|---|---|---|
| | Multi-dimensional ID coordinate space. | Circular NodeID space. | Mesh network | Mesh network | XOR metric for distance between points in the key space |
| **Parameters** (in addition to N: number of peers) | $d$:number of dimensions. | | $B$-base of the chosen peer identifier | b-number ofbits ($B = 2b$) usedfor the base ofthe chosenidentifier | $b$-number ofbits ($B = 2b$) of NodeID |
| **Routing hops** | $O(\frac{d}{2}N^{\frac{1}{d}})$ | $O(\frac{1}{2}log_2(N))$ | $O(log_B(N))$ | $O(\frac{1}{b}log_2(N))$ | $O(log_b(N))$ |
| **Routing state** | O(2d) | $O(2\ log_2(N))$ | $O(log_B(N))$ | $O(\frac{1}{b}(2^b-1)log_2(N))$ | $O(b\ log_b(N))$ |
| **Join** | $O(\frac{d}{2}N^{\frac{1}{d}})$ | $O(log_2^2(N))$ | $O(log_B(N))$ | $O(log_{2^b}(N))$ | $O(log_b(N))$ |
| **Leave** | O(2d) | $O(log_2^2(N))$ | $O(log_B(N))$ | $O(log_b(N))$ | $O(log_b(N))$ |

Table1: Structured Peer-to-Peer Systems: Performances Analysis

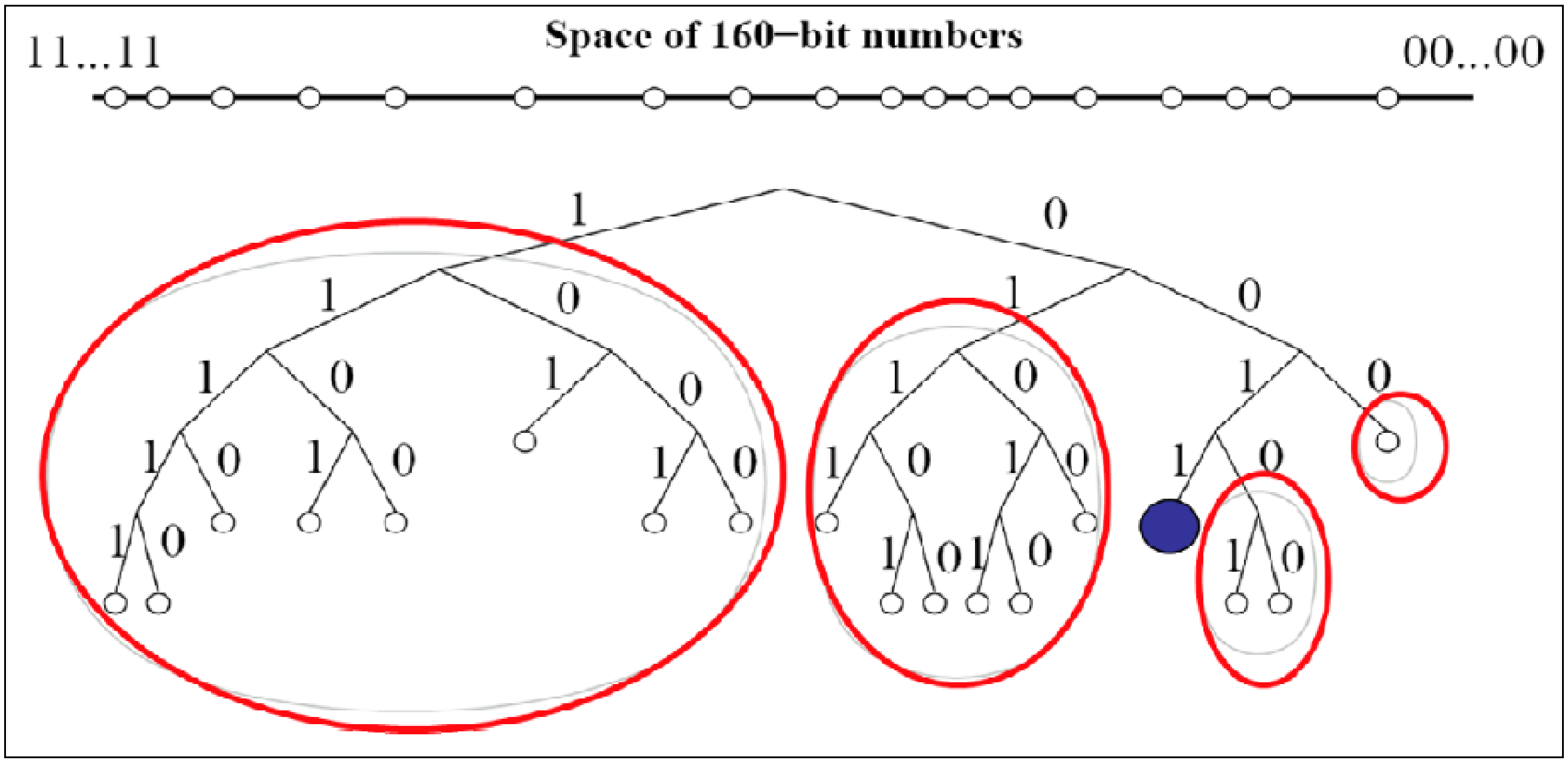

Figure 2: Subtrees of interest for a node 0011

of k is chosen so that any given set of k nodes is unlikely to fail within an hour. This is based on the observation of Gnutella showing that the longer a node is up, the more likely it is to remain up for one more hour. This increases the stability of the routing topology and also prevents good links from being flushed from the routing tables by distributed denial-of-service attacks, as can be the case in other DHT systems.The list is updated whenever a node receives a message. Each k-bucket is kept sorted by last time seen.

- The Kademlia routing protocol consists of the following operations:
- PING probes a peer to check if it is active.
- STORE instructs a peer to store a {key,value} pair for later retrieval.
- FIND_NODE takes a 160-bit ID, and returns {IP address, UDP port, NodeID} triples for the k peers it knows that are closest to the target ID.
- FIND_VALUE is similar to FIND_NODE: it returns {IP address, UDP port, NodeID} triples, except in the case

when a peer receives a STORE for the key, in which case it just returns the stored value.

Next section will discuss the mapping between the operations of the IAX protocol and the Kademlia structures.

## 5. IAX over a Kademlia-based Peer-to-Peer Architecture

This section discusses the main IAX operations affected by the use of Kademlia architecture. Mainly, UA (user agent) registration, UA release, and call flow.

Kademlia architecture is characterized by three parameters: alpha, B, and k. alpha is the number of simultaneous asynchronous requests, also called the degree of parallelism in network calls, the optimal value of alpha is 3 as proved in [19]. B is the size in bits of the keys used to identify peers and store and retrieve data, usually 160 as defined by the Kademlia's author. k is the maximum number of contacts stored in a bucket, usually 20.

IAX provides a facility for one peer to register its address and credentials with another so that callers can reach the registrant. Registration is performed by a peer that sends a REGREQ message to the registrar. If authentication is required, the registrar responds with the REGAUTH message that indicates the types of authentication supported by the registrar. In response, the registrant resends a REGREQ with one of the supported authentications. If accepted, the registrar sends a REGACK message or REGREJ message to indicate a failure. In our case, no registrar is required. The algorithm is as follow:

1. If it does not already have a peerID, it generates one
2. It inserts the value of some known peer into the appropriate bucket as its first contact. If the contact list is empty, it adds the registration server as its first contact.
3. It does a registration for its own peerID by sending a REGREQ message to its first contact. This step will populate other peers' k-buckets with its peerID, and will populate the joining peer's k-buckets with the peers in the path between it and the first contact peer.
4. Every peer receiving a REGREQ reply by a REGACK if registration is accepted or REGREJ if it fails.
5. It populates its routing table by the parameters of the peers who has sent a REGACK message.
6. It refreshes all buckets further away than the k-bucket the first contact falls in. This refresh is just a lookup using an IAX PING message of a random key that is within that k-bucket range.
7. A peer receiving a PING message replies with an IAX PONG message. A local peer end the registration process by sending back an IAX ACK message.

Given a specific peerID, the peer runs the following algorithm to get the the k closest peers to a given key. Each peer is stored as a tuple (peerID, IP address).The algorithm FIND_CALLEES for looking up k peers closest to the target ID from the routing table is as follow:

1. It selects alpha contacts from the k-bucket closest to the bucket of the key being searched on. If fewer than alpha contacts are found, contacts are selected from other buckets. The contact closest to the target key is noted *closestNeighbor*. The first alpha contacts are stored in a list noted *searchlist*.
2. It sends parallel, asynchronous FIND_CALLEES message to the alpha contacts in the *searchlist*.
3. Each contact should return k tuples. If a peer fails to reply, it is removed from *searchlist*.
4. It fills the *searchlist* with contacts from the received replies. These are those closest to the target. From the *searchlist*, it selects other alpha contactsthat they have not already been contacted.
5. It updates *closestNeighbo*r. If *closestNeighbo*r is unchanged, then the initiating peer sends a FIND_CALLEES to each of the k closest peers that it has not already queried.
6. It loops until either no peer in the sets returned is closer than the *closestNeighbo*ror k probed and known to be active contacts has been accumulated.

Given a specific peerID, the algorithm FIND_CALLEE for looking up for a peerID (and its parameters such as the IP address) of a given IAX peer address (peerX@serverY.com) follows the same steps of the algorithm FIND_CALLEES. If a peer receives a FIND_CALLEE message and specified peerID is present, it returns its parameters. Otherwise, it returns k tuples as in step 3.

In order to set up a call between peer A and peer B. Peer A send FIND_CALLEE over the overlay network. If B is live, its IP address should be returned to A. Then, A sends an IAX NEW message to B and a normal IAX call flow is applied.

When a peer wants to leave/disconnect, it sends a REGREL message to all contacts of its routing table. Each peer that receives REGREL replies with a REGACK message and forward the message to alpha contacts. The status of the disconnected peer will be changed to offline but not removed from the buckets unless a new peer has a favorite status. A peer that remains offline more than a specified time (24 hours for example) will be removed from the buckets. Offline peers do not participate in any IAX algorithm.

In order to guarantee a load balancing between peers, the routing table should be updated according to the number of peers in the overlay. When the k-bucket becomes full, it will be split. The split occurs if the range of peers in the k-bucket spans peerID. If a peer should be inserted in a local bucket, it means that it shares the longest common prefix with the local peerID in the routing table and it is near the local peerID. Thus, it is a good peer and it should be stored. On the other hand, if the right sub-tree of the minimum sub-tree containing the Local peerID and the target peerID contains at least k peers, it means that there are at least k peers closer than the new peer, and then the coming peer is of less importance, and can be discarded. This is to guarantee that the network knows about all peers in the closest region.

## 6. Conclusion

In this paper, we have proposed an IAX-based peer-to-peer VoIP architecture; an optimized architecture that takes advantage of the benefits of the IAX protocol and those of the peer-to-peer distribution model. The proposed architecture relies on the use of the IAX protocol over a Kademlia peer-to-peer structure. The design of the proposed architecture involves a mapping between the operations of the IAX protocol and the Kademlia structures. In the presented work, we have focused on the mapping between the main operations of the IAX protocol (UA registration, UA release, and call flow) and the Kademlia structure. A future work will consider a total mapping between the IAX operations and Kademlia peer-to-peer architecture, as well as a performance evaluation study of the proposed approach.

**Amor Lazzez** is currently an Assistant Professor of Computer and Information Science at Taif University, Kingdom of Saudi Arabia. He received the Engineering diploma with honors from the high school of computer sciences (ENSI), Tunisia, in June 1998, the Master degree in Telecommunication from the high school of communication (Sup'Com), Tunisia, in November 2002, and the Ph.D. degree in information and communications technologies form the high school of communication, Tunisia, in November 2007. Dr. Lazzez is a researcher at the Digital Security research unit, Sup'Com. His primary areas of research include design and analysis of architectures and protocols for optical networks.

**Ouissem Ben Fredj** received the BE degree in computer science from the University Manar II, Tunisia in 2002. He obtained the MS in computer science from University of Henri Poincare, France in 2003. He Obtained the PhD degree in computer science from University of Val d'Essonnes, France in 2007. He is currently an assistant professor of computer science at Taif University, Saudi Arabia. His research interests include Distributed Systems and network security.

**Thabet Slimani** graduated at the University of Tunis (Tunisia Republic) and defended PhD. thesis with title "New approaches for semantic Association Extraction and Analysis". He has been working as an assistant Professor at the Department of Computer Science, Taif University. He is a member of Larodec Lab (Tunis University). His interests include semantic Web, data mining and web service. He is author of more than 20 scientific publications.